\documentclass[preprint,prb,nobibnotes,showkeys,preprintnumbers,amsmath,amssymb]{revtex4}
\usepackage{graphicx}
\usepackage{pst-all}
\usepackage{dcolumn}
\usepackage{rotating}
\usepackage{bm}
\usepackage{natbib}
\usepackage[dvips]{epsfig}

\newcommand{\ie}{i.\,e.\ }

\begin{document}
\title{Linear mass density of vertically suspended heavy springs.
}
\author{Cyril Belardinelli}
\email[Electronic mail: ]{cyril.belardinelli@hslu.ch}
\affiliation{Lucerne University of Applied Sciences and Arts, Technikumstrasse 21, CH-6048 Horw, Switzerland}
\date{\today}
\begin{abstract}
In the present article the linear mass density $\lambda$ for vertically suspended heavy springs is calculated for two different cases. First for a spring of invariable length suspended at the top and fixed at the bottom of the spring, then for a hanging heavy spring with an additional load. Both cases are solved by minimizing the total energy $E[\lambda]$ which is a sum of potential energy and energy due to the deformation. 
\end{abstract}
\maketitle
\section{\label{intro}Introduction}
Hooke's law for ideal springs is a recurring subject in elementary physics courses. In most textbook problems the mass of the spring itself is neglected. An interesting question arises when one asks for the static deformation of the spring when it is suspended vertically with or without a mass hanging at the free end. Due to its own weight and the loaded mass the spring deforms in a non-uniform way giving rise to a non-constant mass density function $\lambda(x)$. The issue of heavy springs has already been treated rather intensely by various authors.\cite{cushing:1984, mak:1987,ruby:2000, essen:2010, galloni:1979, serna:2010}
However, an explicit formula for the linear mass density of suspended heavy springs does not seem to occur in the literature. The issues presented in this article are nicely solved by the standard methods of variational calculus.\cite{courant:43, goldstein:1980}
\section{\label{fixed_ends}Suspended spring with fixed ends}
In this section we consider an ideal spring of mass \textit{m}, natural length $l_{0}$ and stiffness $\textit{k}$. The spring is vertically suspended where both ends are fixed at a distance $\textit{l}>\textit{l}_{0}$.
\begin{figure}[h!]
\centering
\includegraphics[width=7cm]{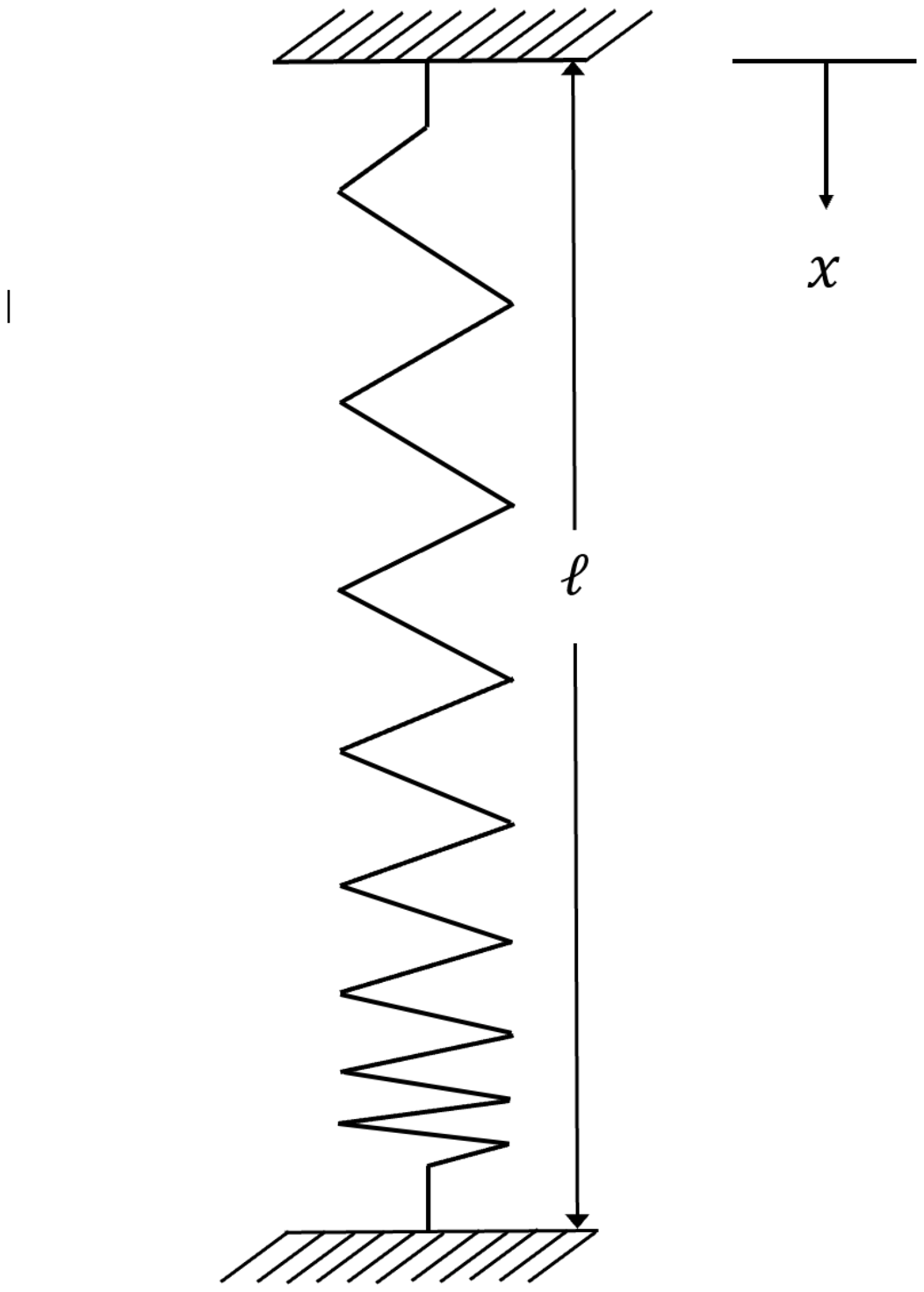}
\caption{vertically suspended ideal spring of stiffness $k$, natural length $l_{0}$ and mass $m$ with fixed ends. The spring has in its stretched form a length of $l$. The spring's linear mass density in its undeformed state is $\lambda_{0}=\frac{m}{l_{0}}$. In its suspended position, the spring's linear mass density $\lambda(x)$ increases from top to bottom due to its weight. }
\end{figure}
\\To find the linear mass density $\lambda(x)$ one minimizes the energy functional $E_{\text{tot}}[\lambda]$ which is given by the sum of the potential energy $E_{pot}$ and deformation energy $E_d$:
\begin{equation}
\label{E_tot}
E_{\text{tot}}[\lambda]=E_{\text{pot}}[\lambda]+E_d[\lambda]
\end{equation}
The potential energy formulated as a functional of $\lambda(x)$ is given by:
\begin{equation}
\label{E_pot}
E_{\text{pot}}[\lambda]=-g\int_{0}^{\textit{l}}x\lambda(x)dx
\end{equation}
The energy functional $E_d[\lambda]$ due to deformation is given by:
\begin{equation}
E_{d}[\lambda]=\frac{kl_{0}}{2}\int_{0}^{\textit{l}}\left(\frac{\lambda_{0}}{\lambda(x)}-2+\frac{\lambda(x)}{\lambda_{0}}\right)dx 
\end{equation}
One has to minimize Eq.~(\ref{E_tot}) under the constraints of invariable mass and non-negativity of $\lambda(x)$:
\begin{equation}
\label{constraint}
m=\int_{0}^{\textit{l}}\lambda(x)dx \quad \text{and} \quad \lambda(x)>0 \quad \forall x\in [0,l]
\end{equation}
In order to take account of the constraints one introduces a Lagrange-multiplier $\mu$ with which the functional reads as follows:
\begin{equation}
\label{F}
F[\lambda]=E_{\text{tot}}[\lambda]-\mu\left[m-\int_{0}^{\textit{l}}\lambda(x)dx\right]
\end{equation} 
Variation of the functional Eq.~(\ref{F}) leads to the following Euler-Lagrange-equation.\cite{courant:43,goldstein:1980}:
\begin{equation}
\frac{kl_{0}\lambda_{0}}{2}\left(\frac{1}{\lambda_{0}^2}-\frac{1}{\lambda^{2}(x)}\right)-gx+\mu=0
\end{equation}
From the latter equation one deduces ($\lambda_{0}l_{0}=m$):
\begin{equation}
\label{lambda_x}
\lambda(x)=\frac{\lambda_{0}}{\sqrt{1+\frac{2\mu m}{kl^2_{0}}-\frac{2mg}{kl^2_{0}}x}}
\end{equation} 
The Lagrange-multiplier  $\mu$ in Eq.~(\ref{F}) is fixed by the constraint Eq.~(\ref{constraint}). A short calculation yields:
\begin{equation}
\mu=\frac{k}{2m}(l^2-l^2_{0})+\frac{mg^2}{8k}+\frac{gl}{2}
\end{equation}
By inserting $\mu$ into Eq.~(\ref{lambda_x}) one gets the final form for the linear mass density $\lambda(x)$. For further calculations an expression in terms of dimensionless variables is convenient. The reduced mass density $\tilde{\lambda}:=\frac{\lambda}{\lambda_{0}}$ reads then:
\begin{equation}
\label{mass_density_fixed_ends}
\tilde{\lambda}(\tilde{x})=\frac{1}{\sqrt{(\frac{\alpha}{2}+\tilde{l})^2-2\alpha\tilde{x}}} ,\quad 0<\tilde{x}<\tilde{l}
\end{equation}
where the dimensionless parameters read:
\begin{equation}
\alpha:=\frac{mg}{kl_{0}}  \quad \tilde{x}:=\frac{x}{l_{0}} \quad \tilde{l}:=\frac{l}{l_{0}}
\end{equation}
\begin{figure}[h!]
\centering
\includegraphics[width=8cm, angle=-90]{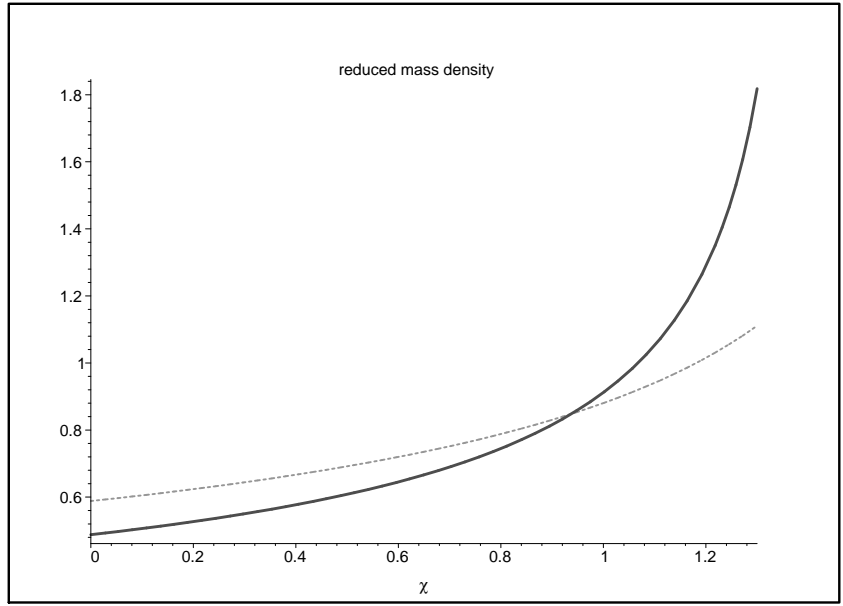}
\caption{reduced mass density $\tilde{\lambda}(\tilde{x})$ of Eq.~(\ref{mass_density_fixed_ends}) for $\alpha=0.8$ (dotted curve)
and $\alpha=1.5$ (solid). For both plots: $\tilde{l}=1.3$}
\end{figure}
\section{\label{free_end}heavy spring suspended at the top end with load}
\begin{figure}[h!]
\label{figure}
\centering
\includegraphics[width=7cm]{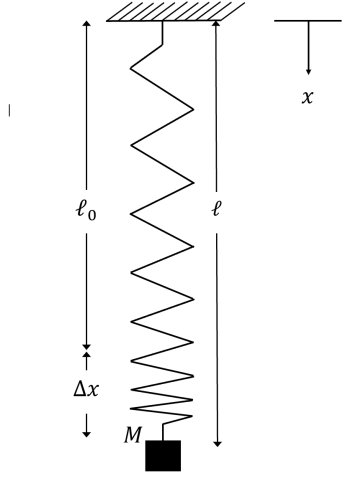}
\caption{free hanging ideal spring of stiffness $k$, natural length $l_{0}$ and mass $m$. The elongation $\Delta x$ is due to the spring's own weight and additional load $M$.}
\end{figure}
In this section we calculate the linear mass density $\lambda(x)$ for a static spring of mass $m$ and stiffness $k$ which is suspended only at the top end. The spring is loaded with an additional mass $M$ at the bottom end.
The calculation can be done in almost the same manner as in the previous section. The calculation starts again with the same functional Eq.~(\ref{lambda_x}) which has to be minimized. To fix the Lagrange-multiplier $\mu$ one can use the fact that at the bottom end the spring is only deformed by the additional load $M$. It is a simple task to derive the spring's mass density $\lambda(l)$ at the bottom end, as shown in the following.\\
$dx$ denotes an infinitesimal intervall of the suspended spring at the bottom end. The corresponding natural (undeformed) length of $dx$ is denoted by $dx_{0}$. 
By Hooke's law one has:
\begin{equation}
\label{lambda_bottom}
k^{\prime}(dx-dx_{0})=Mg
\end{equation}
where $k^{\prime}$ denotes the stiffness of the spring of infinitesimal length $dx_{0}$ which is given by:
\begin{equation}
k^{\prime}=k\frac{l_{0}}{dx_{0}}
\end{equation} 
Inserting the latter expression into Eq.~(\ref{lambda_bottom}) one gets:
\begin{equation}
kl_{0}\left(\frac{dx}{dx_{0}}-1\right)=Mg
\end{equation}
By using the relation: $\frac{\lambda_{0}}{\lambda(l)}=\frac{dx}{dx_{0}}$ one arrives at:
\begin{equation}
kl_{0}\left(\frac{\lambda_{0}}{\lambda(l)}-1\right)=Mg
\end{equation}
Solving for $\lambda(l)$ yields:
\begin{equation}
\lambda(l)=\frac{\lambda_{0}}{1+\frac{Mg}{kl_{0}}}
\end{equation}
For an unloaded spring (\ie $M=0$) the density at the bottom is, as expected, identical to $\lambda_{0}$ since the spring does not experience any weight at the bottom.
Then by Eq.~(\ref{lambda_x}) one gets the Lagrange-multiplier $\mu$:
\begin{equation}
\mu=gl+\frac{M}{m}gl_{0}+\frac{M^2g^2}{2mk}
\end{equation}
At this stage one can determine the total length $l$ of the spring by imposing once more the constraint of invariable mass:
\begin{equation}
\label{constraint_2}
\int_{0}^{l}\lambda(x)dx=m
\end{equation}
Rewriting the latter integral in terms of the usual dimensionless variables (by adding the dimensionless parameter $ \beta:=\frac{Mg}{kl_{0}} $) one gets:
\begin{equation}
\int_{0}^{\tilde{l}}{\frac{d\tilde{x}}{\sqrt{(1+\beta)^2+2\alpha(\tilde{l}-\tilde{x})}}}=1
\end{equation}
Evaluation of the integral and solving for the total relative length $\tilde{l}$ gives:
\begin{equation}
\label{total_length}
\tilde{l}=1+\frac{\alpha}{2}+\beta
\end{equation}
or respectively:
\begin{equation}
l=l_{0}+\frac{mg}{2k}+\frac{Mg}{k}
\end{equation}
For an unloaded spring (\ie $M=0$) the elongation $\Delta x$ due to its own weight is given by $\Delta x=\frac{mg}{2k}$ which is a 
well known result.\cite{mak:1987, galloni:1979}
By Eq.~(\ref{total_length}) one gets immediately an expression for the reduced mass density:
\begin{equation}
\label{mass_density_free_end}
\tilde{\lambda}(x)=\frac{1}{\sqrt{(1+\alpha+\beta)^2-2\alpha \tilde{x}}},\quad 0<\tilde{x}<\tilde{l}
\end{equation}
\begin{figure}[h!]
\centering
\includegraphics[width=8cm, angle=-90]{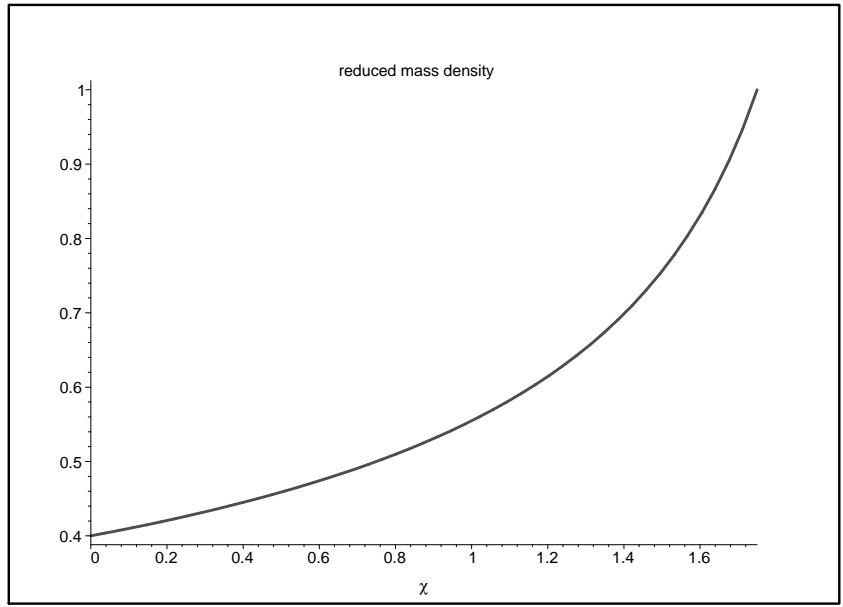}
\caption{Typical plot of the reduced mass density $\tilde{\lambda}(\tilde{x})$ given by Eq.~(\ref{mass_density_free_end}) for no load ($M=0$) and $\alpha=1.5$. At the bottom end of the spring (\ie $x=l_{0}+\frac{mg}{2k}$) one has a reduced mass density which assumes the value $\tilde{\lambda}=\frac{\lambda(l)}{\lambda_{0}}=1$ as shown by the plot.}
\end{figure}
\section{\label{E_d_E_pot}relative change of $\Delta E_{\text{pot}}$ vs. $\Delta E_{\text{d}}$}
The distribution of mass along the suspended spring is fixed by the condition of minimizing total energy. It is interesting to determine the ratio $ \frac{\Delta E_{pot}}{\Delta E_{d}}$ of relative change in potential energy and deformation energy induced by the spring's deformation. 
\subsection{\label{E_d_E_pot_fixed_ends} spring with fixed ends}
The relative change in potential energy is given by:
\begin{equation}
\Delta E_{\text{pot}}=-\int_{0}^{l}x\lambda(x)dx+\frac{1}{2}mgl_{0}=-mgl_{0}\int_{0}^{\tilde{l}}\frac{\tilde{x} d\tilde{x}}{\sqrt{(\frac{\alpha}{2}+\tilde{l})^2-2\alpha\tilde{x}}}+\frac{1}{2}mgl_{0}
\end{equation} 
The latter integral can be evaluated easily, one gets then
\begin{equation}
\Delta E_{\text{pot}}=-mgl_{0}\frac{\alpha}{12}
\end{equation}
From the latter result one concludes that the lowering of the center of mass is given by:
\begin{equation}
\Delta x=\frac{\alpha}{12}l_{0}=\frac{mg}{12k}
\end{equation}
On the other hand, the relative change in the deformation energy is given by the following expression:
\begin{equation}
\Delta E_{d}=\frac{1}{2}kl_{0}\int_{0}^{l}\left(\frac{\lambda_{0}}{\lambda(x)}-2+\frac{\lambda(x)}{\lambda_{0}}\right)dx-\frac{1}{2}kl_{0}^2(\tilde{l}-1)^2
\end{equation}
In dimensionless form the later expression becomes:
\begin{equation}
\Delta E_{d}=\frac{1}{2}kl_{0}^{2}\int_{0}^{\tilde{l}}\left[\sqrt{\left(\frac{\alpha}{2}+\tilde{l}\right)^2-2\alpha \tilde{x}}-2+\left(\sqrt{\left(\frac{\alpha}{2}+\tilde{l}\right)^2-2\alpha \tilde{x}}\right)^{-1}\right]dx-\frac{1}{2}kl_{0}^2(\tilde{l}-1)^2
\end{equation}
Evaluation of the latter expression yields:
\begin{equation}
\Delta E_{d}=\frac{1}{24}kl_{0}^{2}\alpha^2
\end{equation}
We conclude therefore:
\begin{equation}
\frac{\Delta E_{\text{pot}}}{\Delta E_{d}}=2
\end{equation}
\subsection{\label{E_d_E_pot_free_spring} spring suspended at top end}
To calculate the change in potential energy $\Delta E_{\text{pot}}$ one determines the center of mass (CM) of the suspended spring. It is given by the integral:
\begin{equation}
CM=\frac{1}{l_{0}}\int_{0}^{1+\frac{mg}{2k}}\frac{x dx}{\sqrt{(1+\frac{mg}{l_{0}k})^{2}-2\frac{mg}{kl_{0}}x}}
\end{equation}
Once again, in dimensionless form one has:
\begin{equation}
\begin{split}
CM&=l_{0}\int_{0}^{1+\frac{\alpha}{2}}\frac{\tilde{x} d\tilde{x}}{\sqrt{(1+\alpha)^{2}-2\alpha \tilde{x}}}=\left(\frac{1}{2}+\frac{\alpha}{3}\right)l_{0}\\
&=\frac{l_{0}}{2}+\frac{mg}{3k}
\end{split}
\end{equation}
Due to the self weight of the spring the center of mass lowers by an amount of $\frac{mg}{3k}$ which is a known result.\cite{serna:2010} With this result one determines immediately $\Delta E_{\text{pot}}$:
\begin{equation}
\Delta E_{\text{pot}}=-\frac{1}{3}k\left(\frac{mg}{k}\right)^2
\end{equation}
The change in deformation energy is readily evaluated. One has:
\begin{equation}
\Delta E_{d}=\frac{1}{2}kl_{0}^{2}\int_{0}^{1+\frac{\alpha}{2}}\left[\sqrt{\left(1+\alpha\right)^2-2\alpha \tilde{x}}-2+\left(\sqrt{\left(1+\alpha\right)^2-2\alpha \tilde{x}}\right)^{-1}\right]dx-\frac{1}{2}kl_{0}^2(\tilde{l}-1)^2
\end{equation}
Evaluation gives:
\begin{equation}
\Delta E_{\text{d}}=\frac{1}{6}kl_{0}^{2}\alpha^2=\frac{2}{3}k\left(\frac{mg}{2k}\right)^2
\end{equation}
Interestingly, one gets the same ratio:
\begin{equation}
\frac{{\Delta E_{\text{pot}}}}{\Delta E_{d}}=2
\end{equation}
\newpage
\appendix*
\bibliography{heavy_spring}
\bibliographystyle{plain}
\end{document}